\documentclass[article, aps,nofootinbib]{revtex4}
\input epsf
\usepackage{graphics}
\usepackage{amsmath}
\usepackage{color}
\usepackage{dcolumn}

\usepackage{graphicx}

\def\be{\begin{equation}}
\def\ee{\end{equation}}
\def\ba{\begin{eqnarray}}
\def\ea{\end{eqnarray}}

\newcommand{\beqa}{\begin{eqnarray}}
\newcommand{\eeqa}{\end{eqnarray}}

\newcommand{\bn}{\hat{\bf n}}
\newcommand{\bl}{\hat{\bf l}}
\newcommand{\beq}{\begin{equation}}
\newcommand{\eeq}{\end{equation}}
\newcommand{\bfl}{{\mathbf{l}}}

\newcommand{\bfL}{{\mathbf{L}}}
\newcommand{\bfLp}{{\mathbf{L^{\prime}}}}
\newcommand{\bflp}{{\mathbf{l^{\prime}}}}

\newcommand{\intl}[1]{\int {d^2 l_{#1} \over (2\pi)^2}}
\newcommand{\intlp}[1]{\int {d^2 l_{#1}' \over (2\pi)^2}}

\newlength{\tskip}
\newlength{\colwidth}

\begin{document}
\title{Probing Primordial Magnetism with Off-Diagonal Correlators 
of CMB Polarization}
\author{
Amit Yadav$^{1,2}$,
Levon Pogosian$^{3}$, 
Tanmay Vachaspati$^{4}$
}

\affiliation{
$^1$Institute for Advanced Study, Princeton, NJ 08540, USA\\
$^2$ Department of Physics, University of California at San Diego, La Jolla, CA 92093 \\
$^3$Department of Physics, Simon Fraser University, Burnaby, BC, V5A 1S6,
Canada \\
$^4$Physics Department, Arizona State University, Tempe, AZ 85287, USA  \\
}

\begin{abstract}
Primordial magnetic fields (PMF) can create polarization $B$-modes in the 
cosmic microwave background (CMB) through Faraday rotation (FR), leading 
to non-trivial 2-point and 4-point correlators of the CMB temperature 
and polarization.
We discuss the detectability of primordial magnetic fields using
different correlators and evaluate their relative merits.
We have fully accounted for the contamination by weak lensing, which
contributes to the variance, but whose contribution to the 4-point correlations
is orthogonal to that of FR. 
We show that a Planck-like experiment can detect scale-invariant PMF of 
nG strength using the FR diagnostic at $90$GHz, while realistic future 
experiments at the same frequency can detect $10^{-10}\, {\rm G}$. 
Utilizing multiple frequencies will improve on these prospects,
making FR of CMB a powerful probe of scale-invariant PMF.
\end{abstract}

\maketitle

\section{Introduction}
Magnetic fields are prevalent in the cosmic structures around us,
in galaxies  $B\sim 1\, \mu {\rm G}$ with coherence length 
$\lambda \sim 1 \, {\rm kpc}$, in galaxy clusters  
$B\sim 1-10\, \mu {\rm G}$ with coherence length 
$\lambda \sim 10-100\, {\rm kpc}$, and in objects at high redshifts 
$z\sim 2$ with magnetic field 
$B\sim 10 \, \mu {\rm G}$~\cite{RevModPhys.74.775}. 
Recently there have also been claims 
of a lower bound on the inter-galactic magnetic field,
$B> 10^{-15} {\rm G}$ 
\cite{Neronov_Vovk_2010_science,Dolag:2010ni,Taylor:2011bn,Vovk:2011aa}, 
and perhaps a measurement $\sim 10^{-15}\, {\rm G}$ 
\cite{Ando:2010rb}, 
based on the absence of GeV $\gamma$-ray emission in the cascade initiated 
by TeV $\gamma$-rays. The claim is under debate as it has been argued that plasma instabilities could also explain the nonobservation of GeV photons~\cite{Broderick:2011av}, though a counterargument that supports the initial claim has been presented in Ref.~\cite{Miniati_Elyiv_2012}
It is possible that these magnetic fields have a common origin from a 
``seed'' magnetic field imprinted in the early universe 
(see \cite{Grasso_Rubinstein_2001} for a review). 
Magnetic fields may be generated at cosmic phase transitions
~\cite{Vachaspati:1991nm,
Cornwall:1997ms,Vachaspati:2001nb,GarciaBellido:2003wd,Copi:2008he,
Vachaspati:2008pi,Ng:2010mt,Chu:2011}
and through specially engineered inflationary 
mechanisms~\cite{Turner:1987bw,Ratra:1991bn}.

Detection of primordial magnetic fields (PMF) can lead to important insights
into fundamental physics and the early
universe. Currently, there are upper limits on the strength of PMF
from big-bang nucleosynthesis 
(BBN)~\cite{Matese_OConnell_69,1996PhRvD..54.7207K, Cheng:1996yi,Grasso:1996kk,Kawasaki:2012va}
and the Cosmic Microwave Background (CMB) temperature 
anisotropies~\cite{Finelli:2008xh,Paoletti:2008ck,
2009PhRvD..80b3009K,Paoletti:2010rx}, including their non-Gaussian statistics,
such as bispectrum and trispectrum \cite{Seshadri:2009sy,Trivedi:2011vt}.
Metric fluctuations induced by PMF are intrinsically
non-Gaussian because the stress-energy is quadratic in the magnetic field
strength ${\bf B}$ and thus non-Gaussian distributed even if ${\bf B}$ itself
is Gaussian. In this paper we study the detectability of PMF through a
different observational window, namely, the Faraday Rotation (FR) signal they
induce in the CMB polarization~\cite{Kosowsky:1996yc}. The polarization of the
CMB field can be studied in terms of the parity even $E$ and parity odd
$B$-modes~
\cite{Kamionkowski:1996zd,SZ97,1997PhRvD..55.7368K}. 
FR converts some of the primordial $E$-modes into $B$-modes, thus providing a
contribution
to the $B$-mode power spectrum along with the weak gravitational lensing and
primordial sources, like the actively sought inflationary gravity waves
\cite{Crittenden:1993wm,Kamionkowski:1996zd,SZ97} or cosmic strings
\cite{Seljak:1997ii}. 

Importantly, spatially dependent FR also couples off-diagonal CMB modes,
effectively producing additional non-Gaussian signatures in the CMB
polarization. The FR induced parity odd correlations $\langle TB \rangle$ and $\langle
EB\rangle$ must vanish in a statistically isotropic 
universe\footnote{This statement is specific
to the FR induced parity-odd two-point correlations, which are quadratic in the
magnetic field 
strength ${\bf B}$. Metric fluctuations sourced by
magnetic stress-energy can produce non-zero $\langle TB \rangle$ and $\langle
EB\rangle$, which are quartic in ${\bf B}$, if the net helicity in
the magnetic field is non-zero \cite{Kahniashvili:2005xe,Kunze:2011bp}.}, where
the ensemble average is
over many realizations of the stochastic magnetic field. However,  
{\it a particular realization} of the FR distortion field that generates
a $B$-mode from the primordial $E$-mode will correlate the
respective Legendre coefficients $E_{lm}$ and $B_{l'm'}$. In fact, as shown in
\cite{Kamionkowski:2008fp}, it is possible to reconstruct the distortion
$\alpha(\bn)$ at a given point $\bn$ on the sky from specially constructed
linear combinations of products $E_{lm}B_{l'm'}$. The additional correlations
induced by FR also manifest themselves as connected 4-point functions of the
CMB, which, in turn, provide a measurement of the distortion spectrum
$C_L^{\alpha \alpha}$
\cite{Yadav_etal_09,2009PhRvD..80b3510G}.

In this paper we discuss the detectability of primordial magnetic fields 
through
estimators based on 4-point correlations $\langle EBEB \rangle$ and $\langle
TBTB \rangle$, as well as the 2-point function $\langle BB \rangle$. In
particular, we determine which of the estimators has the highest signal to
noise for several types of magnetic field spectra and for a range of
experimental sensitivities. We demonstrate that FR will be a very promising
diagnostic of primordial magnetic fields. In particular, future generation of
sub-orbital or space-based CMB polarization experiments will be able to detect
scale-invariant magnetic fields as weak as $10^{-10}$G based on the measurement
at $90$ GHz
frequency. Measurements at multiple frequencies can further
significantly improve on these prospects.
 
\section{Faraday Rotation from Magnetic Fields}
\label{sec:PNGB}

Magnetic fields at CMB decoupling will rotate the polarization vector by an angle
\begin{equation}
\alpha = \frac{3}{{16 \pi^2 e}} \lambda_0^2 
\int \dot{\tau} \ {\bf B} \cdot d {\bf l} \ ,
\label{theta2}
\end{equation}
where $\dot{\tau} \equiv n_e \sigma_T a$ is the differential 
optical depth, $n_e$ is the line of sight free electron density, 
$\sigma_T$ is the Thomson scattering cross-section, $a$ is the scale factor, 
$\lambda_0$ is the observed wavelength of the radiation, ${\bf B}$ is
the ``comoving'' magnetic field, and $d{\bf l}$ is the comoving
length element along the photon trajectory.
We are using Gaussian natural units 
with $\hbar=c=1$, and the integration limits are from the initial 
to the final position of the photon.

Statistically homogeneous and isotropic primordial seed magnetic 
fields can be generated in the early universe 
during phase transitions \cite{Vachaspati:1991nm,Cornwall:1997ms,
Vachaspati:2001nb,GarciaBellido:2003wd,Copi:2008he,Vachaspati:2008pi,
Ng:2010mt,Chu:2011}, and are described in terms of a two-point 
correlation function in Fourier space 
\be
\langle b_i ({\bf k} ) b_j ({\bf k}' ) \rangle =
(2\pi )^3 \delta^{(3)}({\bf k} + {\bf k}' )
[ (\delta_{ij} - {\hat k}_i {\hat k}_j) S(k) +
i \varepsilon_{ijl} {\hat k}_l A(k) ] \ ,
\label{bcorr}
\ee
where $S(k)$ and $A(k)$, the symmetric and anti-symmetric magnetic
power spectra, are real functions of $k=|{\bf k}|$. The function 
$A(k)$ quantifies the amount of magnetic helicity which plays a 
crucial role in determining the coherence scale and the magnitude of 
magnetic fields as they evolve from an earlier epoch until decoupling. 
However, only $S(k)$ appears in the two-point correlation function of 
the FR angle, which determines the CMB observables evaluated in this 
paper. As in \cite{Pogosian:2011qv}, we introduce 
the dimensionless ``FR power spectrum'' defined as 
\begin{eqnarray}
\Delta^2_M(k) &\equiv& k^3 S(k) \left( \frac{3 \lambda_0^2}{16\pi^2 e}
\right)^2 
= \begin{cases}
 \Delta^2_0  \left ( \frac{k}{k_I} \right )^{2n} &\mbox{$0<k<k_I$}\\ 
 \Delta^2_0  \left ( \frac{k}{k_I} \right )^{2n'} &\mbox{$k_I<k<k_{\rm diss}$}\\
 0 &\mbox{$k>k_{\rm diss}$}
 \end{cases} \ ,
\label{eq:deltaB}
\end{eqnarray}
where
\begin{equation}
\nonumber
\Delta^2_0 \equiv \frac{9n}{16\pi e^2 \kappa} \rho_\gamma {\lambda_0^4}  
              ~  \Omega_{B\gamma} \approx 1.1\times 10^4 ~ \frac{\Omega_{B\gamma}}{\kappa} \times 
 \left ( \frac{2n}{5} \right ) \left ( \frac{90~{\rm GHz}}{\nu_0} \right )^4 \ .
\end{equation}
This form of the symmetric magnetic spectrum is based on the numerical simulations
of causal magnetic field evolution in~\cite{Jedamzik:2010cy}. In the above,
$\rho_{\gamma}$ is the comoving photon energy density, 
$\Omega_{B\gamma}$ is the magnetic energy density relative to 
the photon 
energy density, $k_{\rm diss}$ is a dissipation scale above which magnetic 
fields dissipate, $k_I$ is an intermediate inertial scale, and
\begin{eqnarray}
\kappa=1+\frac{n}{n'} \left\{ \left( \frac{k_{\rm diss}}{k_I} \right)^{2n'}-1 
\right\}\,.
\label{eq:kappa}
\end{eqnarray}
All variables, unless explicitly stated, are in comoving coordinates. 
The exponents $n=5/2$ and $n'=3/2$ correspond to causal magnetic 
fields~\cite{Jedamzik:2010cy}, and $n \approx n' \approx 0$ are 
expected for magnetic fields generated in an inflationary 
scenario~\cite{Turner:1987bw,Ratra:1991bn}. The dissipation scale, $k_{\rm diss}$ 
is not an independent parameter and should, in principle, be dependent on the
amplitude and the shape of the magnetic fields spectrum. As in \cite{Pogosian:2011qv}, we
assume that $k_{\rm diss}$ is determined 
by damping into Alfven waves \cite{Jedamzik:1996wp,Kahniashvili:2010wm} and can be 
related to $B_{\rm eff}$ as
\be
{k_{\rm diss} \over 1{\rm Mpc}^{-1}} \approx
          1.4 \ h^{1/2} \left( 10^{-7} {\rm Gauss} \over B_{\rm eff} \right) \ ,
\label{kIBeff}
\ee
where $B_{\rm eff}$ is defined as the effective homogeneous field strength
that would have the same total magnetic energy density. It is related
to $\Omega_{B \gamma}$ via \cite{Pogosian:2011qv}
\be
B_{\rm eff}= 3.25 \times 10^{-6} \sqrt{\Omega_{B \gamma}} \ {\rm Gauss} \ .
\label{beff-omega}
\ee
Subsequently, the dissipation scale can be expressed in terms of $\Omega_{B \gamma}$ as 
\be
k_{\rm diss} \approx 0.43 \sqrt{10^{-2} h \over \Omega_{B\gamma}}
\ {\rm Mpc}^{-1} \ .
\label{kIOmegaB}
\ee
One should be aware of the very approximate nature of the relations (\ref{kIBeff})
and (\ref{kIOmegaB}). They are based on the analysis in Ref.~\cite{Jedamzik:1996wp} 
where small perturbations 
on top of a homogeneous magnetic field were treated. To extend this analysis to a 
stochastic magnetic field with little power on long wavelengths, Ref.~\cite{Kahniashvili:2010wm} 
introduced a smoothing procedure and split the spectrum into a ``homogeneous'' 
part and a ``perturbations'' part. 
It is not clear to us if this procedure is valid for an arbitrary spectrum,
$S(k)$. Instead, we will use Eq.~(\ref{kIBeff}) as an approximate expression 
for the dissipation scale. We note that this relation also imposes an upper
bound on $k_I$, since $k_I$ cannot be greater than $k_{\rm diss}$.

In practice, both scales ($k_I$ and $k_{\rm diss}$) that appear in the
definition of
the magnetic spectrum (\ref{eq:deltaB}) are likely to be smaller than the
resolution
scale of a realistic CMB experiment. This means that only the large scale tail
of the spectrum, {\it i.e.} the $0<k<k_I$ range, will be relevant for 
calculating the shapes of
the CMB correlation functions on the observable scales. The existence of the
intermediate range $k_I < k < k_{\rm diss}$ and the exponent $n'$ will affect 
the inferred constraints on $\Omega_{B\gamma}$ only through an overall 
rescaling. For this reason, in the calculation of the CMB correlations, we will
set $k_I=k_{\rm diss}$, with $k_{\rm diss}$ given by (\ref{kIOmegaB}),  and
work with a single power law spectrum
\be
\tilde{\Delta}^2_M(k) 
= \begin{cases}
 \tilde{\Delta}^2_0  \left ( \frac{k}{k_{\rm diss}} \right )^{2n} &\mbox{$0<k<k_{\rm diss}$}\\ 
 0 &\mbox{$k>k_{\rm diss}$}
 \end{cases} \ ,
\label{eq:singlePB}
\ee
where
\be
\tilde{\Delta}^2_0 = 1.1\times 10^4 ~ \tilde{\Omega}_{B\gamma} \times 
 \left ( \frac{2n}{5} \right ) \left ( \frac{90~{\rm GHz}}{\nu_0} \right )^4
\label{eq:tilde-delta0} 
\ee
The bound we obtain on $\tilde{\Omega}_{B\gamma}$ can then be easily converted
to the bound on $\Omega_{B\gamma}$ via 
\be
\Omega_{B\gamma} = \tilde{\Omega}_{B\gamma} \kappa \left( {k_I \over k_{\rm diss}} \right)^{2n} \ .
\label{eq:conversion}
\ee

We calculate the FR power spectrum as \cite{Pogosian:2011qv}
\begin{eqnarray}
C_L^{\alpha \alpha} = {2\over \pi } \int {dk \over k} \tilde{\Delta}^2_M(k)
\left[ 
{L \over 2L+1} ({\cal T}_{L-1}(k))^2 + 
{L+1 \over 2L+1} ({\cal T}_{L+1}(k))^2 - ({\cal T}_L^{(1)}(k))^2
 \right] \ ,
\label{c_ell} 
\end{eqnarray}
where ${\cal T}_L(k)$ are transfer functions that are independent of the magnetic field:
\ba
\nonumber
{\cal T}_L(k) &\equiv&  
\int_{\eta_*}^{\eta_0} d\eta  ~ {\dot \tau}(\eta) j_L(k(\eta_0-\eta)) \\
{\cal T}_L^{(1)}(k) &\equiv&  
\int_{\eta_*}^{\eta_0} d\eta  ~ {\dot \tau}(\eta) j_L'(k(\eta_0-\eta)) \ .
\label{eq:transfer}
\ea
Here $\eta_*$ is the epoch at which the visibility function is maximum, $j_L$
are the spherical Bessel functions, and ${\dot \tau}$ can be easily obtained 
numerically using public codes such as CMBFAST \cite{cmbfast} or CAMB \cite{Lewis:1999bs}.

\section{The mode coupling estimator and the $B$-mode spectrum}  
                           
FR will rotate the CMB polarization fields generated at last scattering. This introduces coupling
between different CMB modes which can, in fact, be used to reconstruct the rotation angle map from 
the observed CMB polarization maps \cite{Kamionkowski:2008fp,2009PhRvD..80b3510G,Gluscevic:2012me,YSZ09}.
Let $\tilde{T}(\bn)$, $\tilde{Q}(\bn)$ and  $\tilde{U}(\bn)$ be the un-rotated 
CMB temperature field and the two linear polarization Stokes parameters 
at angular position $\bn$. The temperature is not affected by FR, except
for the depolarization effect \cite{Harari:1996ac}
which would appear as a next order correction and can be ignored in our
analysis. Under a rotation of the polarization by an angle $\alpha(\bn)$, 
the two Stokes parameters transform like a spin two field. Thus, the 
observed fields are
\begin{equation}
(Q(\bn) \pm iU(\bn))
=(\tilde{Q}(\bn) \pm \tilde{U}(\bn)) \exp(\pm 2i\alpha(\bn)) \ .
\label{RotationTransformationOfStokes}
\end{equation}
 The observed Stokes parameters can be further combined to form
parity-even ($E$-mode) and parity-odd ($B$-mode) combinations
which, in the flat sky 
approximation~\cite{Kamionkowski:1996zd,SZ97,1997PhRvD..55.7368K}
are defined by
\begin{eqnarray}
 \left[ E\pm i B \right] (\bfl) &=&
        \int  d \bn\, [Q(\bn)\pm i U(\bn)] e^{\mp 2i\varphi_{\bf l}} e^{-i \bl \cdot \bn}\,,
\label{EBFields}
\end{eqnarray}
where $\varphi_{\bfl}=\cos^{-1}(\hat {\bf n} \cdot \hat \bfl)$. 
The relevant\footnote{We assume that the unrotated CMB temperature 
and polarization fluctuations are Gaussian distributed.} ensemble averages 
of the un-rotated CMB fields can be encapsulated in
\begin{equation}
\left\langle \tilde{x}(\bfl)\right\rangle = 0, \qquad
\left\langle \tilde{x}^\star(\bfl)\tilde{x}^{\prime}(\bfl^{\prime})\right\rangle
=
(2\pi)^2 \delta(\bfl -\bfl^{\prime})\tilde{C}_{\bfl}^{xx^{\prime}}\,,
\end{equation}
where ${\tilde x}, {\tilde x}^{\prime}$ run over the $T, E$, or $B$ fields, and $\tilde{C}_{\bfl }^{xx^{\prime}}$ 
are the un-rotated CMB power spectra.

We would like to isolate the effect of FR on the various correlators
that describe the CMB. In particular, gravitational lensing causes
CMB distortions that can interfere with the computation of the FR
effect\footnote{In this paper we do not utilize the frequency 
dependence of the FR as a tool to differentiate it from other
effects.}. 
The change due to FR is the 
difference between lensed rotated fields and lensed un-rotated fields 
$\tilde{T},\tilde{E},\tilde{B}$. The change in the CMB fields 
$\delta x(l) = x(l) -\tilde{x}(l)$ due to FR is
\begin{eqnarray}
\delta T(\bfl) &=& 0 \,,\\
\delta B(\bfl)   &=& 2\intlp{}
\Big[ \tilde E(\bfl') \cos 2\varphi_{\bfl'\bfl}
     -\tilde B(\bfl') \sin 2\varphi_{\bfl'\bfl} \Big] \alpha(\bfL)
,\nonumber\\
\delta E(\bfl)      &=& -2\intlp{}
\Big[ \tilde B(\bfl') \cos 2\varphi_{\bfl'\bfl}
     +\tilde E(\bfl') \sin 2\varphi_{\bfl'\bfl}\Big] \alpha(\bfL),\nonumber
\end{eqnarray}
where $\bfL=\bfl - \bflp$, and
$\varphi_{\bfl \bflp}= \varphi_\bfl - \varphi_\bflp$.
Thus, due to FR, a mode of wavevector $\bfL$ mixes the polarization
modes of wavevectors $\bfl$ and $\bf{l^{\prime}}=\bfl -\bfL$. 

Let us take the ensemble average over multiple realizations of the 
unrotated CMB fields, while assuming a fixed $\alpha$ field. Then, 
for the rotated variables $x\ne x'$, one can write
\begin{equation}
\langle x^\star(\bfl) x'(\bflp) \rangle_{\rm CMB} =  
                     f_{xx'}(\bfl,\bflp) \alpha(\bfL)\,,
\label{BasicDifference}
\end{equation}
where $f_{TB}=2 {\tilde C}_{l}^{T E}\cos 2\varphi_{\bfl\bfl'}$,  and $f_{E B}=2 [\tilde C_{l}^{E E}-\tilde C_{l'}^{B B}]
\cos 2 \varphi_{\bfl\bfl'}$. 
Namely, a single realization of a random rotation field $\alpha$ will
induce parity-odd correlations of types $TB$ and $EB$ that
are linearly proportional to the FR angle. If we also average 
over an ensemble of magnetic fields, the above 
two point functions will vanish for $x\ne x'$, since the 
expectation value of the magnetic field is zero.

The mode-coupling rotation (\ref{BasicDifference}) imprinted in the 
CMB by FR implies that one can build estimators for reconstructing the 
FR field from the observed CMB. 
Following Ref.~\cite{2002ApJ...574..566H,Kamionkowski:2008fp,
Yadav_etal_09,2009PhRvD..80b3510G,YSZ09}, we can define 
an unbiased estimator $\hat{\alpha}_{xx'}(\bfL)$ for $\alpha(\bfL)$, 
where $x \ne x'$,
by taking quadratic combinations of different polarization modes 
weighted by 
a factor $F_{xx'}(\bfl_1,\bfl_2)$:
\begin{eqnarray}
\hat \alpha_{xx'}({\bfL})& =&  N^{xx'}_L \intl{1}
\, x(\bfl_1) x'(\bfL-\bfl_1) 
F_{xx'}(\bfl_1,\bfL-\bfl_1)\,,
\label{eqn:estimator}
\end{eqnarray}
where $\bfL=\bfl_2  -\bfl_1$, 
and the normalization
\begin{eqnarray}
N^{xx'}_L = \Bigg[ \intl{1} f_{xx'}(\bfl_1,\bfL-\bfl_1)
F_{xx'}(\bfl_1,\bfL-\bfl_1) \Bigg]^{-1} \,,
\label{eq:noise}
\end{eqnarray}
is chosen to make the estimator unbiased, i.e. $\langle
\hat\alpha(\bfL)\rangle_{\text{CMB}}=\alpha(\bfL)$. 
The fields $x(\bfl)$ can be obtained from the map of an experiment, while the
CMB power spectrum of un-rotated but lensed fields can be computed from 
publicly available
Boltzmann codes like CMBfast \cite{cmbfast} and CAMB \cite{Lewis:1999bs}. 
The weights $F_{xx'}$ are determined by minimizing the variance subject to 
the normalization constraint.
For $x x'=T B$ and $EB$ the minimization yields
\begin{equation}
F_{xx'}(\bfl_1,\bfl_2) = \frac{f_{xx'}(\bfl_1,\bfl_2)}
                                {C_{l_1}^{xx} C_{l_2}^{x'x'}}\,,
\end{equation}
where $C_{l_2}^{xx}$ and $C_{l_2}^{x'x'}$ are the observed power spectra
including the effects of both the signal and the instrument
 \begin{eqnarray}
C_{l}^{xx}=\tilde C_{l}^{xx}+\Delta^2_x e^{l^{2}\theta^2_{\text{FWHM}}/(8\ln{2})}
\end{eqnarray}
where $\Delta_x$ is the detector noise and  $\theta_{\text{FWHM}}$ is the full-width half-maximum (FWHM) resolution 
of the Gaussian beam \cite{Eisenstein:1998hr}.

The variance of the estimator can be calculated as
\begin{eqnarray}
\text{Var}(\hat \alpha_{xx'}(\bfL))&=&
  \langle {\hat \alpha}_{xx^\prime}(\bfL) 
 {\hat \alpha}^\star_{xx^{\prime}}(\bfL^{\prime})\rangle \nonumber \\
 &=& N^{xx'}_L N^{xx'}_{L'} \intl{1} \intl{2}
\, \langle x(\bfl_1) x'(\bfL-\bfl_1)x(\bfl_2) x'(\bfL '-\bfl_2)\rangle 
\, F_{xx'}(\bfl_1,\bfL-\bfl_1)F_{xx'}(\bfl_2,\bfL '-\bfl_2) \nonumber\\
 &=&  (2\pi)^2 \delta(\bfL - \bfLp) 
        \{ C^{\alpha \alpha}_L +N^{xx^\prime}_L\}\,.
\label{eqn:variance}
\end{eqnarray}
In the last line, the
first term is the desired FR power spectrum and the second term is 
the noise -- also referred to as the Gaussian noise -- in the reconstruction 
of FR. Note that the noise turns out 
to coincide with the normalization of the minimum variance estimator 
as given by Eq.~(\ref{eq:noise})~\cite{Kamionkowski:2008fp,2009PhRvD..80b3510G,Gluscevic:2012me,YSZ09},
and is independent of 
FR. In principle, there are higher order noise terms 
(referred to as non-Gaussian noise terms) which depend on FR, however, 
these terms are sub-dominant in comparison to the Gaussian noise.
The signal-to-noise for detecting a spatially varying 
FR angle $\alpha(\bn)$ using the estimator (\ref{eqn:estimator}) 
is given by~\cite{Kamionkowski:2008fp, Yadav_etal_09}
\begin{eqnarray}
\left(\frac{S}{N}\right)^2_{xx'}&=& \sum^{l_{max}}_{l=2}\frac{f_{\rm sky}}{2} 
    (2l+1) \left  ( \frac{C^{\alpha \alpha}_l}{N^{xx'}_l}\right )^2 \,,
\label{eq:s2n}
\end{eqnarray}
where $C^{\alpha \alpha}_l$ is the rotation angle power spectrum, 
and we ignore the contribution of $C^{\alpha \alpha}_l$ to the 
variance (\ref{eqn:variance}) as it is negligible compared to $N^{xx'}_l$.

We also consider the case when $x=x'=B$, and the
the resultant $BB$ correlation is quadratic in the rotation field. 
Averaging over the ensemble of magnetic fields gives
the FR induced CMB $B$-mode power spectrum: 
\begin{eqnarray}
C^{BB}_L= 4\int \frac{d^2{\bf l'}}{(2\pi)^2} 
 C^{\alpha \alpha}_{l'} C^{EE}_{l''}
       \cos^2[2\varphi_{\bfl'' \bfL}]
    \biggr |_{\bfl''=\bfL-\bfl'} 
\label{eq:clbb}
\end{eqnarray} 
The signal-to-noise in this case (accounting for the 
$B$-modes from weak lensing but assuming no
contribution from inflationary gravity waves) is given by 
\begin{eqnarray}
\left(\frac{S}{N}\right)^2_{BB}&=& \sum^{l_{max}}_{l=2}\frac{f_{\rm sky}}{2} 
    (2l+1) \left  ( \frac{C^{BB}_l}{{\cal N}^{BB}_l}\right )^2 \,,
\label{eq:s2n-bb}
\end{eqnarray}
where ${\cal N}_l^{BB}=C_l^{BB,\text{lensing}}+C_l^{BB,\text{noise}}$ and 
the noise power spectrum is $C_l^{BB,\text{noise}} = 
\Delta^2_P \exp({l^2\Theta^2_{\text{FWHM}}/8 \ln 2 })$, 
where $\Delta_P$ is the instrument noise for polarization.

One might wonder if gravitational lensing, which also generates 
off-diagonal correlations in the CMB with a leading order contribution to the
trispectrum, might bias the mode coupling FR estimator (\ref{eqn:estimator}). 
However, it has been shown that 
the effects of gravitational lensing and rotation are orthogonal and 
hence lensing does not bias estimates of rotation
\cite{Yadav_etal_09,2009PhRvD..80b3510G}. 
Lensing does, however, increase the variance of the estimator, 
i.e. the Gaussian noise given by Eq.~(\ref{eq:noise}) is enhanced. 
For example, the noise of the $EB$ estimator, $N^{EB}_\ell$, 
depends on the observed CMB $E$ and $B$ mode power spectra, 
and is an integration over $(C^{EE}_\ell+{\cal N}^{EE}_\ell) 
(C^{BB}_\ell+{\cal N}^{BB}_\ell)$. Further, the $B$-modes due to 
lensing scale as white noise up to $l\sim 1000$,
and correspond to a noise level of $5 \mu$K-arcmin. 
Therefore, the increase in the variance is sub-dominant for experiments 
with $\Delta_P > 5 \mu$K-arcmin. However, for small instrumental 
noise, $\Delta_P \leq 5 \mu$K-arcmin, lensing $B$-modes become 
important, saturating the variance to $\sim10^{-6} ~{\rm  deg}^2$ 
even for an ideal experiment.

In the next section, we compare detectability of the FR
signal in upcoming and future CMB maps with the 
mode-coupling estimators (\ref{eqn:estimator}) vs. 
using the $B$-mode spectrum (\ref{eq:clbb}). Interestingly, we find
that one method can out-perform the other depending on
the value of the magnetic spectral index.

\section{Detectability of Primordial Magnetic Fields}
\label{sec:detectability}

To forecast the detectability of FR we consider three experimental setups:
a Planck-like satellite \cite{Planck:2006aa} (E1), a ground- or balloon-based
experiment realistically achievable in the next decade (E2), and a future
dedicated CMB polarization satellite (E3).
We will adopt the following parameters for the three experiments:
\begin{itemize}
\item E1: a Planck-like satellite \cite{Planck:2006aa} with noise level $\Delta_P=60\mu K$-arcmin 
and $\Theta_{\rm FWHM}=7'$ 
\item E2: a next decade sub-orbital experiment with $\Delta_P=3.0 \mu K$-arcmin and 
$\Theta_{\rm FWHM}=1'$
\item E3: a CMBPol-like instrument with $\Delta_P=\sqrt{2}\mu K$-arcmin and $\Theta_{\rm FWHM}=4'$, typical of 
the proposed future space-based CMB experiments \cite{Baumann:2008aq,Bouchet:2011ck} 
\end{itemize}
The forecasts directly depend on the fraction of the sky covered by the
experiment, $f_{\rm sky}$,  which is close to unity for E1 and E3, and will be
smaller for E2. We quote our bounds subject to specifying $f_{\rm sky}$ which
only appears under a quartic root in the bounds on $B_{\rm eff}$.

\begin{figure*}[tbp]
\includegraphics[width=110mm,clip,angle=-0]{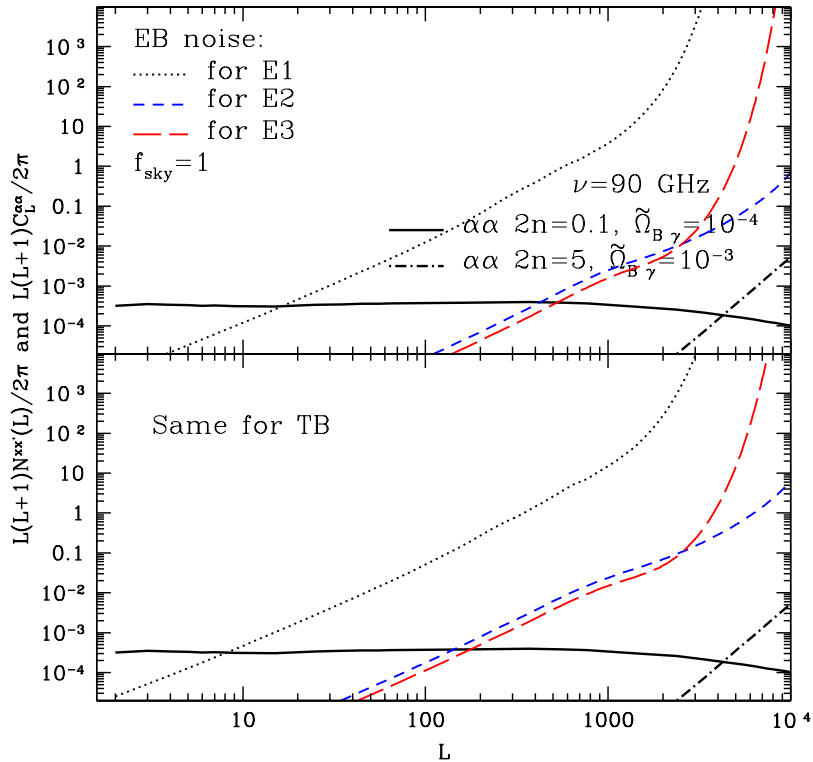}
\caption{The noise $L(L+1)N^{xx'}_L/2\pi$ as given by Eq.~(\ref{eq:noise}) for 
the $EB$ (upper panel) and $TB$ (lower panel) estimators as a function of 
multipole $L$ for three experimental setups (E1, E2, E3). Also plotted is the 
FR power spectrum $L(L+1) C^{\alpha \alpha}_{L}/2\pi$ for the causal magnetic
spectrum with $2n=5$ and for a nearly scale 
invariant spectrum with $2n=0.1$. The amplitude of the
FR spectrum is set by the observed frequency of $\nu_0=90$ GHz and 
$\tilde{\Omega}_{B\gamma}=10^{-3}$  for the causal case ($2n=5$), and 
$\tilde{\Omega}_{B\lambda}= 10^{-4}$ for the nearly scale invariant case ($2n=0.1$). }
\label{fig1}
\end{figure*}

In Fig.~\ref{fig1} we plot $L(L+1)N^{xx'}_L/2\pi$ vs multipole $L$ with 
the noise for the $EB$ and $TB$ estimators given 
by Eq.~(\ref{eq:noise}) for experiments E1, E2 and E3, along with the FR
spectrum $L(L+1) C^{\alpha \alpha}_{L}/2\pi$ for two choices of the magnetic spectral index $2n$. 
The amplitudes of the rotation spectra are normalized to $\tilde{\Omega}_{B\gamma}=10^{-3}$ 
for the causal case ($2n=5$) and 
$\tilde{\Omega}_{B\gamma}= 10^{-4}$ for the nearly scale invariant case ($2n=0.1$), with
$\tilde{\Omega}_{B\gamma}$ related to $\Omega_{B\gamma}$ via 
Eq.~(\ref{eq:conversion}). For all the
estimates in this paper we adopted an observational frequency of $\nu_0=90$
GHz, but one can easily scale the signal to other frequencies using
Eq.~(\ref{eq:tilde-delta0}). For all three experiments under consideration, the $EB$ estimator is more
sensitive than the $TB$, with the noise $N^{xx'}_L$ staying roughly constant up to  $L\sim 1000$. 
Although we do not show  $L=0$ in the plot, these estimators can also be 
used to estimate the detectability of uniform rotation. 

\begin{figure*}[tbp]
\includegraphics[width=110mm,clip,angle=-0]{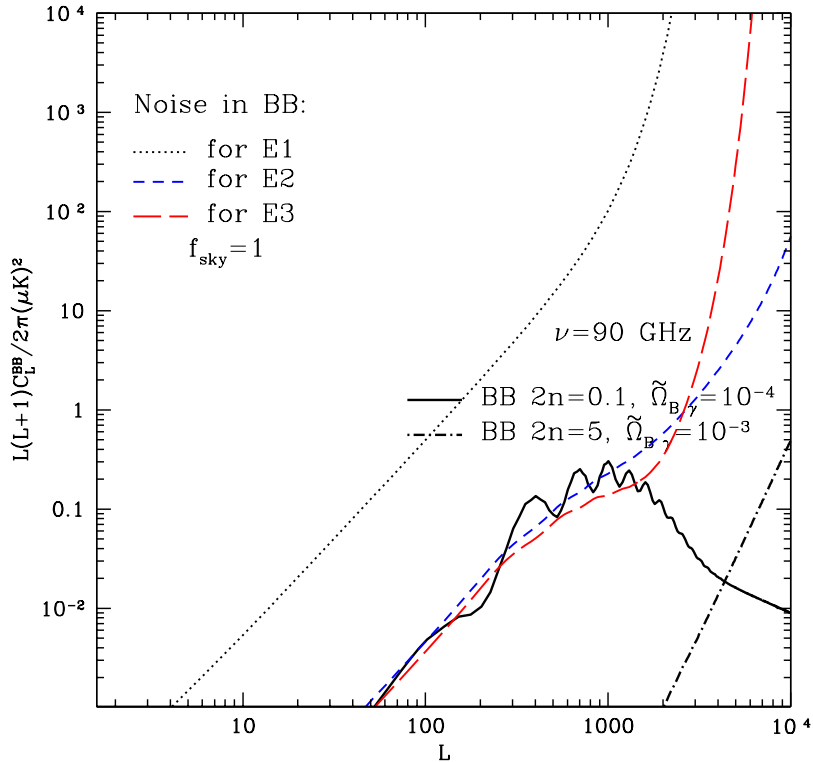}
\caption{The $B$-mode polarization pixel noise, 
$L(L+1)C_L^{BB,\text{noise}}/2\pi$, for 
experiments E1, E2, and E3 as a function of 
multipole $L$. Also shown is the $B$-mode spectrum induced by FR
for a causal magnetic spectrum with $2n=5$ and for a nearly scale 
invariant spectrum with $2n=0.1$. The observed frequency is set to
$\nu_0=90$ GHz, $\tilde{\Omega}_{B\gamma}= 10^{-3}$  for the 
causal case, and $\tilde{\Omega}_{B\gamma}= 10^{-4}$ for the scale 
invariant case. }
\label{fig2}
\end{figure*}

In Fig.~\ref{fig2} we plot the noise contribution to the variance of the $B$-mode
spectrum, along with the FR induced $B$-mode spectrum for
$\tilde{\Omega}_{B\gamma}= 10^{-3}$  for $2n=5$, and $\tilde{\Omega}_{B\gamma}=
10^{-4}$ for $2n=0.1$. In the latter case, the shape of the $B$-mode spectrum is
a close copy of the 
underlying $E$-mode, except that the FR induced spectrum falls off as
$L(L+1)C^{BB}_L \propto L^{2n-1}$ at high $L$
\cite{2005PhRvD..71d3006K,Pogosian:2011qv} compared to the exponential fall off
of the primordial $E$-mode. The $L^{2n-1}$ tail implies a sharply rising spectrum
for the causal case with ($2n=5$), with most of the power concentrated near the
dissipation scale $L_{\rm diss} \sim 10^4\, {\rm Mpc} \, \times k_{\rm diss}$.

\begin{figure*}[tbp]
\includegraphics[width=120mm,clip,angle=-0]{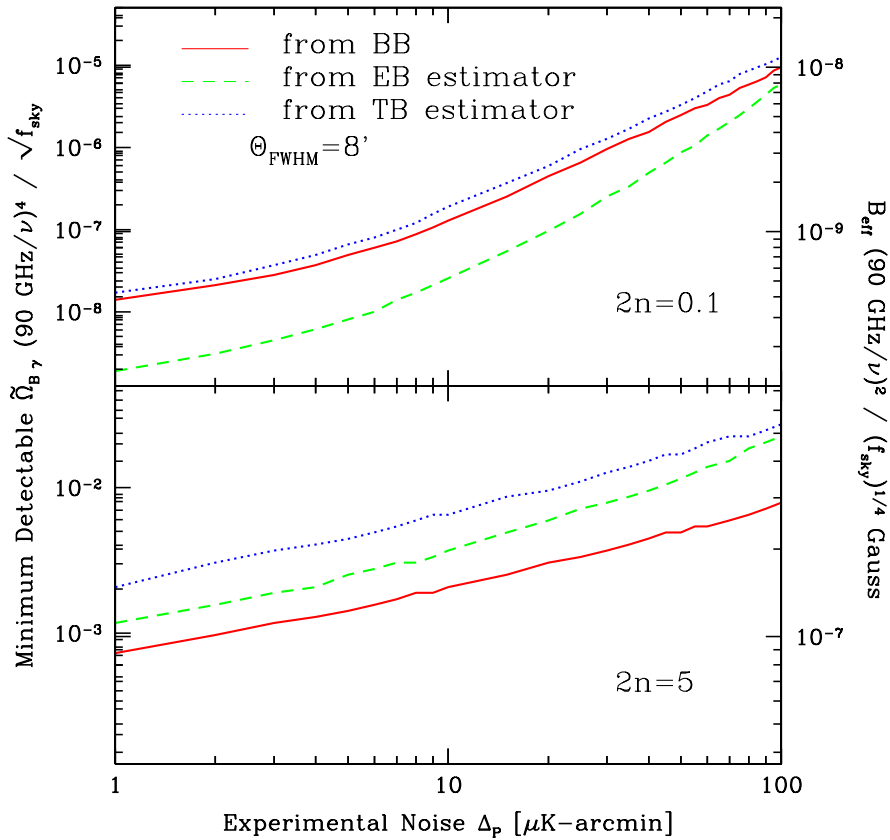}
\caption{Comparison of the three estimators, $BB$ (solid red), $EB$ (dashed green) and $TB$ (dotted blue). Plotted is the minimum detectable magnetic field amplitude $\tilde{\Omega}_{B\gamma}$ as a function of the experimental noise, with $\Theta_{\rm FWHM}=8'$. The upper panel is for the nearly scale invariant case with $2n=0.1$, while the lower panel is for the causal case with $2n=5$.}
\label{fig3}
\end{figure*}

To forecast the minimum detectable magnetic field energy fraction, we define it
as $\tilde{\Omega}_{B\gamma}$ for which $S/N$ in Eqs.~(\ref{eq:s2n}) and
(\ref{eq:s2n-bb}) is unity. Note that $\tilde{\Omega}_{B\gamma}$ determines the
dissipation scale $k_{\rm diss}$ via Eq.~(\ref{kIOmegaB}). We perform this
forecast for each estimator, for different experiments, and for several choices
of $2n$. We restrict the maximum multipole to $L_{\rm max}=10000$. 
Fig.~\ref{fig3} shows the minimum detectable 
$\tilde{\Omega}_{B\gamma}$ as a function of the instrument noise $\Delta_P$ for
a fixed beam $\Theta_{\rm FWHM}=8'$ for two choices of $2n$. As one can see, at
high noise levels ($\Delta_P \gtrsim 100 \mu$K-arcmin), 
the $B$-mode power spectrum tends to either give comparable or better constraints
than the $EB$ estimator. In particular, it is always the better probe of the
causal primordial magnetic 
fields. However, for upcoming polarization sensitive experiments with lower
levels of noise, 
the $EB$ estimator will become almost comparable to the $B$-mode power spectrum
for causal fields, and outperform it when probing scale-invariant FR fields. 

\begin{figure*}[tbp]
\includegraphics[width=120mm,clip,angle=-0]{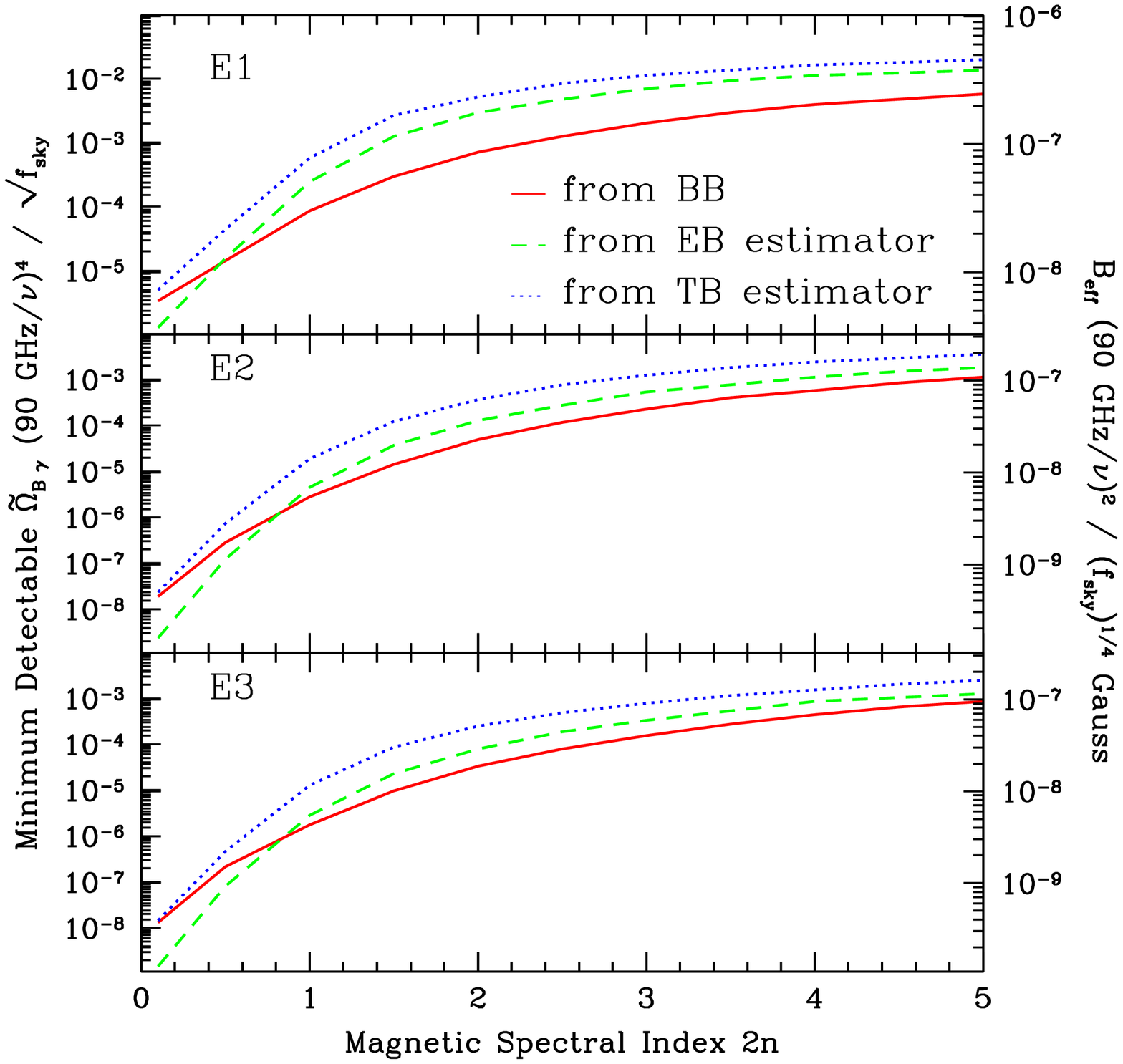}
\caption{The minimum detectable magnetic field amplitude
$\tilde{\Omega}_{B\gamma}$ as a
function of the magnetic spectral index $2n$ for the three estimators, $BB$
(solid
red), $EB$ (dashed green) and $TB$ (dotted blue). The top panel is for E1, the
middle panel is for E2 and the lower panel is for E3.
}
\label{fig4}
\end{figure*}

In Fig.~\ref{fig4} we plot the minimum $\tilde{\Omega}_{B\gamma}$ that will
be detectable by the E1, E2 and E3 experiments depending on the value of the
magnetic spectral index $2n$. We see that for all experiments, the $EB$
estimator 
begins to outperform the $B$-mode spectrum when $2n \lesssim 1$, while the $TB$
estimator is always the third best.

The relative strengths of the three estimators, demonstrated in Figs.~\ref{fig3} and \ref{fig4},
can be understood as follows.
Generally, the $EB$ and $TB$ estimators have a larger number
of independent modes contributing to the signal than the $B$-mode spectrum. 
Thus, in principle, it is not
surprising if they result in a higher signal to noise. However, whether
that is the case depends on the experimental noise level, and the distribution of power in
the given combination of CMB fields and in the magnetic field.
For a scale-invariant PMF spectrum, the $B$-mode is essentially a copy of the
$E$-mode, with most of the $B$-mode power being on scales where
the $E$-modes are also most prominent. This results in a
strong correlation between $E$ and $B$ 
for scale-invariant fields.
In the case of the $TB$ correlation, the underlying $T$ and $E$ ($B$ is obtained
by a scale-invariant rotation of $E$) fields peak 
on rather different scales. Namely, $T$ peaks at $\ell \sim 200$ while $E$ peaks at $\ell \sim 1000$. 
In other words, the intrinsic correlation between $T$ and $E$ is already suboptimal,
translating into a lesser correlation between $T$ and $B$.
Thus, for experiments with a sufficiently low noise $\Delta_P$, such as E1, E2 and E3 considered in this paper, the $EB$ estimator performs better than $TB$ for scale-invariant fields. This would not necessarily remain true if polarization measurements had a significantly higher experimental noise.

For the blue causal spectra, the FR power is concentrated on very small scales, far
away from the scales at which any of the unrotated CMB fields have significant power. 
This means that the $B$-modes in
the observable range are obtained either by the rotation of $E$-modes 
far away from their peak power scale, or by a rotation of peak $E$-mode by
a negligible angle. This means that $E$ and $B$ fields peak at very different scales,
with their correlation being close to zero over the observable scales. In this case,
we see that the $B$-mode spectrum, i.e. the $BB$ correlation, has the highest signal to noise.

When interpreting the forecasted bounds on the magnetic field energy fraction or the
effective magnetic field strength in Figs.~\ref{fig3} and \ref{fig4}, several points must 
be kept in mind: 
\begin{enumerate}
\item The constraints are on $\tilde{\Omega}_{B\gamma}$, obtained after
setting $k_I=k_{\rm diss}$, with the dissipation scale determined from Eq.~(\ref{kIOmegaB}).
For scale-invariant fields there is no difference between
$\tilde{\Omega}_{B\gamma}$ and $\Omega_{B\gamma}$, since the factor relating them
in Eq.~(\ref{eq:conversion}) goes to unity when $2n \rightarrow 0$. Also, for scale-invariant fields,
the effective field $B_{\rm eff}$ defined via Eq.~(\ref{beff-omega}) is the same as the commonly used
$B_\lambda$, which is the field smoothed on a given scale $\lambda$. Thus, our forecasts of
the minimum detectable $B_{\rm eff}$ for scale-invariant fields can be directly compared to 
most other bounds in the literature. 
\item For causal fields, the bound on $\tilde{\Omega}_{B\gamma}$
will generally overestimate the magnetic energy fraction, since it assumes that the spectrum
will keep rising at the same steep rate ($2n=5$) all the way to the dissipation scale, which is much
smaller than the smallest scale directly probed by CMB experiments. Simulations \cite{Jedamzik:2010cy} 
suggest that the spectrum must become less steep, with $2n'=3$ in the range 
$k_I <k< k_{\rm diss}$, implying a smaller net magnetic energy fraction $\Omega_{B\gamma}$.
Since the value of $k_I$ is not well-known at this point, we chose to quote our bounds in terms
of $\tilde{\Omega}_{B\gamma}$, while keeping in mind that bounds on $\Omega_{B\gamma}$ for
causal fields will generally be tighter. 
\item Strictly speaking, our bounds are on the fraction in magnetic fields at
the time when 
the initial conditions for the transfer functions (\ref{eq:transfer}) were set,
which is a time close to last scattering.
While it is expected that the magnetic fields are effectively frozen-in between the
BBN and last scattering,
with a relatively slow time evolution of the dissipation scale, this is still
an approximation.
\item The bounds are based on using a single frequency band. Using several
bands will improve the constraints.
\end{enumerate}

Generally, CMB is not very sensitive to magnetic fields with blue spectra
because most of the anisotropies are concentrated on very small scales. This is
what Figs.~\ref{fig3} and \ref{fig4} are showing too. However, looking for the
FR signatures at many frequencies can potentially improve the existing CMB
bounds on causal fields by a large factor. We leave this question as a topic
for future exploration.

In the case of scale-invariant fields, current bounds on the magnetic field
strength from WMAP are at a level of a few nG
\cite{Finelli:2008xh,Paoletti:2008ck,
Paoletti:2010rx,Seshadri:2009sy,Trivedi:2011vt}.
These bounds are based on the anisotropies induced by the metric fluctuations
sourced by magnetic fields, and ignore the FR effect. 
In Refs.~\cite{2009PhRvD..80b3009K,Pogosian:2011qv} the WMAP bound using
FR was obtained at the $10^{-7}\, {\rm G}$ level.
As one can see from
Fig.~\ref{fig4}, Planck (E1) can almost match today's bounds for scale
invariant ($n=0$) fields using the $EB$
estimator at only one frequency, while future probes, such as E2 and E3,
can improve the bounds by an order of magnitude!
This suggests that the mode coupling estimators of FR will be a very powerful,
if not the most powerful, direct probe of scale-invariant magnetic fields at
the time of last scattering.

\section{Summary}
\label{sec:summary}

A primordial magnetic field present at and just after last scattering
will Faraday-rotate the plane of polarization of 
the CMB photons. The FR will create 
$B$-mode polarization even in the absence of primordial sources,
such as gravity waves. In addition to a $B$-mode autocorrelation, 
FR also couples different modes of the CMB fields, 
generating specific off-diagonal correlators. Estimators
of the polarization rotation angle,  such as (\ref{eqn:estimator}), 
can utilize these non-Gaussian features to reconstruct the FR map. 
One can construct four such estimators containing products of 
two CMB fields, one of which contains polarization: $TE, EE, TB,$ and $EB$. 
Of these four, the first two receive a large contribution to their variance 
from the usual scalar adiabatic Gaussian perturbations which makes it harder 
to find the FR signal. In this paper, we have considered the last two and 
found that the $EB$ estimator has the highest signal-to-noise.

For causal magnetic fields, which tend to have very blue power spectra, the
$B$-mode power spectrum has a higher signal-to-noise due to reasons explained
in the previous section. 
However, the $EB$ estimator performs better for scale-invariant
fields. In addition, there are certain advantages in using mode-coupling 
based estimators, such as $EB$, over the traditional $B$-modes power spectrum.
For instance, there are other sources that can generate $B$-modes, such as weak
lensing, patchy reionization, inflationary tensor perturbations or cosmic strings, and,
in fact, metric perturbations induced by the magnetic fields.
Hence, one has to separate the FR induced $B$-modes 
either by using the frequency dependence of FR or features in the $B$-mode spectrum. 
Interestingly, while patchy reionization and lensing also generate off-diagonal
correlations in CMB, their contributions are ``orthogonal'' to the features imprinted 
by FR~\cite{YSZ09}. Hence, mode-coupling estimators do not suffer from contamination from 
other contributions. This means that a larger part of the information
in the frequency dependence can be used  for systematic checks and 
to separate from other foreground contamination. In addition, mode-coupling
estimators can be used to reconstruct the map of FR. This FR map can be 
used to cross-correlate with other 
tracers of magnetic fields, including CMB maps and surveys of large 
scale structure. Such correlations studies are useful for systematic 
checks and for increasing the signal-to-noise. 

We have found that a Planck-like experiment at 90GHz can detect
scale-invariant PMF of a 
few nG strength, which is comparable to the $\sim$nG sensitivity 
forecasted for Planck based on information in the CMB temperature anisotropies 
\cite{Paoletti:2010rx}. 
Future CMB experiments will be able detect scale-invariant fields as weak as $10^{-10}$ G
at 90GHz. Thus, FR should
become a leading diagnostic of PMF when analyzing future CMB polarization data.

\acknowledgments
APSY gratefully acknowledges funding support from NASA award number 
NNX08AG40G and NSF grant number AST-0807444. LP is 
supported by a Discovery Grant from the Natural Sciences and Engineering 
Research Council of Canada. TV is supported by the Department 
of Energy at ASU, and is grateful to the Institute for Advanced Study, 
Princeton for hospitality. We thank Soma De for pointing out inconsistencies in the noise plots in Figs 1 and 2 that appeared in an earlier version of the manuscript, that we have subsequently corrected.

\end{document}